\documentstyle[11pt,newpasp,twoside,epsf]{article}
\def\etal\/{{\em et al.}}

\def\edcomment#1{\iffalse\marginpar{\raggedright\sl#1\/}\else\relax\fi}
\reversemarginpar

\begin{document}
\title{Polarization properties of 6.7~GHz methanol masers in NGC6334F}
\author{Simon Ellingsen}
\affil{School of Mathematics and Physics, GPO Box 252-37, 
  University of Tasmania, Hobart 7001, TAS, Australia}

\begin{abstract}
  The Australia Telescope Compact Array (ATCA) has been used to make
  the first full polarization observations of 6.7~GHz methanol masers.
  Linear polarization was detected towards all four sources observed,
  at levels between a few and 10\%, while none of the sources show
  circular polarization stronger than approximately 1.5\%.  Linear
  polarization appears to be more common in the 6.7~GHz methanol maser
  transition than it is for the 12.2~GHz transition, consistent with
  the hypothesis that the 6.7~GHz masers are more saturated.
\end{abstract}

\section{Introduction}

Masers from the 6.7~GHz transition of methanol were first detected
towards star formation regions by Menten (1991).  The 6.7~GHz
transition is the strongest of the class~II methanol masers and has
been detected towards more than 400 sources.

One aspect of class~II methanol masers which has received relatively
little attention to date has been the polarization properties.  The
polarization properties of masers are strongly influenced by the
magnetic properties of the molecule producing the emission.  OH is a
paramagnetic molecule and frequently exhibits very high levels of
circular polarization.  Water is a diamagnetic molecule and
consequently has a much smaller magnetic dipole moment, typically
water masers show modest levels of linear polarization, but no
circular polarization.  Methanol is also a diamagnetic molecule and so
it is expected to have polarization characteristics similar to water
masers.  Koo \etal\/ (1988) made a single dish study of the
polarization of five 12.2~GHz methanol masers.  They detected no linear
polarization stronger than 3\% towards three of these (G188.94+0.89,
Cep~A \& NGC\,7538), linear polarization of a few percent towards
W3(OH) and 6-10\% towards NGC6334F.  The observations reported here
are the first detailed studies of the polarization properties of
6.7~GHz methanol masers.

\section{Observations}

The observations were made on 1999 September 22 and 25 using the ATCA
in the 6A configuration, which yields maximum and minimum baseline
lengths of 337~m and 5939~m respectively.  The ATCA has linearly
polarized feeds and the correlator was configured to record a 4~MHz
bandwidth with 1024 spectral channels for each of the products XX, YY,
XY and YX on all baselines.  This resulting velocity resolution was
approximately 0.21~kms$^{-1}$.  Four strong 6.7~GHz methanol
maser sources were observed, G339.88-1.26, G345.01+1.79, NGC6334F and
G9.62+0.20. Five minute scans on each being interleaved with one minute
scans on nearby phase calibrator sources.  In order to obtain good
parallactic angle coverage additional scans were schedule during
transit for each source.  Bandpass and primary flux calibration was
obtained through a 30~minute observation of PKS1934-638 on each of the
two days of observations.

The data were processed using the {\tt miriad} software package using
standard procedures for ATCA observations.  The data for each of the
two days were editted and calibrated independently and the results
compared to check for consistency prior to averaging.  In the absence
of any systematic errors in the data or calibration, the accuracy to
which polarization can be measured with the ATCA is determined by the
signal to noise ratio in the observations used to determine the
leakage terms.  For these observations a 30~minute scan on PKS1934-638
was used to determine the antenna bandpasses and polarization leakage
corrections.  After averaging all baselines together, the signal to
noise ration for the PKS1934-638 observation was estimated to be
approximately 200:1, suggesting that the accuracy of the polarization 
measurements should be 0.4\%.

\section{Results}

Linear polarization was detected at levels between a few and 10\% in a
number of spectral components, in each of the primary target sources.
Most components exhibit linear polarization at a level of about 5\% or
less, although some features have no linear polarization (within the
observational limits).  For example in NGC6334F there are number of
maser components in the velocity range of -10.0 -- -6.5~kms$^{-1}$
with peak flux densities greater than 50~Jy, which show no linear
polarization stronger than 0.4\%.  The rest of this paper will discuss
only the results for NGC6334F, the results for other sources can be
found in (Ellingsen 2002).

Koo \etal\/ (1988) detected linear polarization in two 12.2~GHz
methanol masers, W3(OH) and NGC6334F.  Only the latter of these is
within the declination range covered by the ATCA and so it is the only
source for which we are able to make a comparison of the polarization
properties of the two transitions.  The first high-resolution
observations of the 12.2 GHz methanol masers in NGC6334F, made with
the Parkes-Tidbinbilla Interferometer (PTI) by Norris \etal\/ (1988)
found that the emission arises from two clusters which are separated
by approximately 3\arcsec (i.e. not resolved by the 4\arcmin\/ beam
used by Koo \etal\/ in their observations).  One of these clusters is
projected on the leading edge of the H$\sc{II}$ region NGC6334F, while
there is no radio continuum emission stronger than a 5$\sigma$ limit
of 4.8 mJy at the location of the other (Ellingsen, Norris \&
McCulloch 1996). Subsequent VLBI observations by Ellingsen \etal\/
(1996) found that at 12.2~GHz the emission at velocities less than
-11.0~kms$^{-1}$ is primarily located in the offset cluster (hereafter
NGC6334F(NW)), while the emission at velocities greater than
-11.0~kms$^{-1}$ is located in NGC6334F.  This suggests that there
should be relatively little blending of emission between the two
clusters which would confuse the position angles reported by Koo
\etal\/ (1988)

The first high resolution observations of the 6.7~GHz methanol masers
in NGC6334F were made by Norris \etal\/ (1993).  At 6.7~GHz the
velocity ranges for the three clusters overlap significantly.  Spectra
for each of the four stokes parameters in each cluster were obtained
from cleaned (but not self-calibrated) image cubes, and are shown in
Fig.~1 It is apparent from Fig.~1 that many of the maser components in
both clusters are significantly linearly polarized.  The strongest
features in each of the stokes V spectra are 0.05\% and 1.4\% of the
total intensity at the same velocity for NGC6334F and NGC6334F(NW)
respectively.  The latter of these is marginally significant and
further investigations are being undertaken to determine if it is
real.

\begin{figure}
  \plottwo{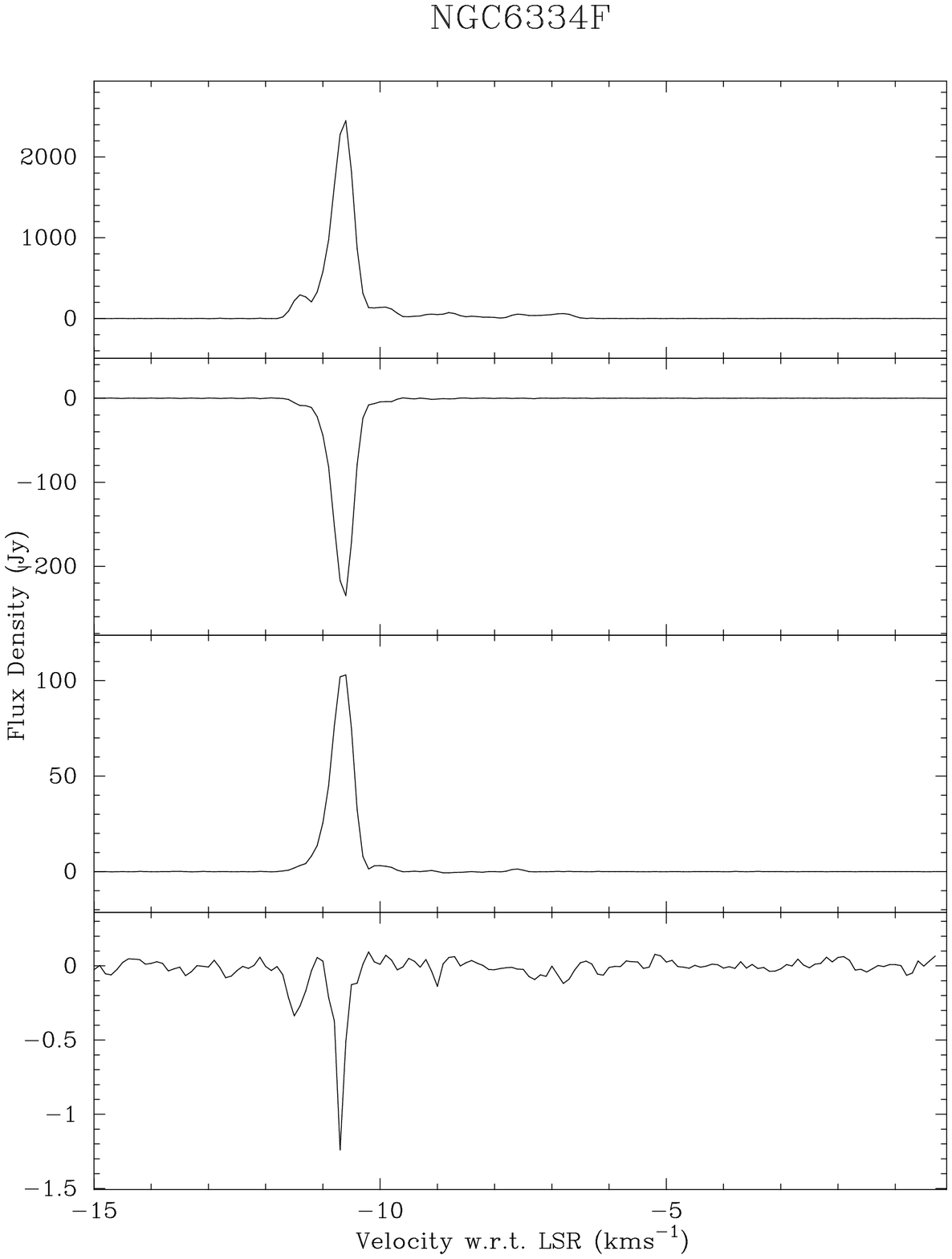}{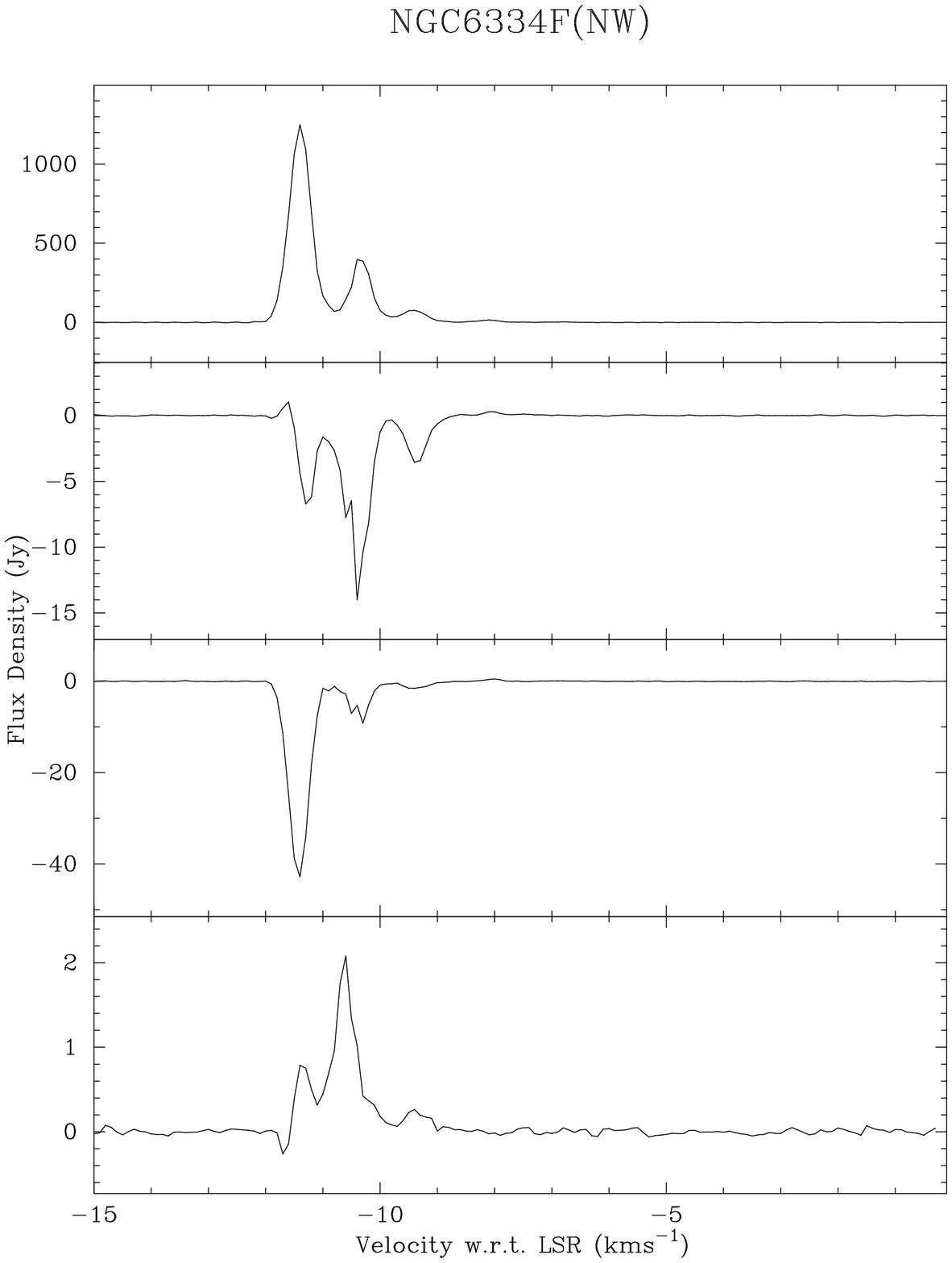}
\caption{The spectrum of the 6.7~GHz methanol maser emission in NGC6334F and 
  NGC6334F(NW) in each of the four stokes parameters I, Q, U and V (from top 
  to bottom in that order)}
\end{figure}

Fig.~2 shows the linear polarization spectrum of the emission from the
two main clusters of 6.7~GHz methanol masers.  In the NGC6334F central
cluster the emission with the highest degree of polarization (10.5\%)
at -10.6~kms$^{-1}$ corresponds to the strongest feature in the total
intensity spectrum.  Interestingly the highest degree of linear
polarization at 12.2 GHz is also 10\% at -10.6~kms$^{-1}$ and the VLBI
observations show that the emission from the two transitions at this
velocity is coincident (Ellingsen \etal\/ 1996).  The position angle
of the emission at 12.2~GHz is quoted by Koo \etal\/ as 45\deg, while
at 6.7~GHz it is 78\deg.  For the NGC6334F(NW) cluster the strongest
component (at -11.4~kms$^{-1}$) also shows the greatest degree of
linear polarization, although in this case it is lower (3.5\%) than
the degree of polarization of the corresponding 12.2~GHz emission
(6\%).  The position angle at 12.2~GHz is 90\deg, while at 6.7~GHz it
is -48\deg.  Since we only have measurements at two frequencies it
isn't possible to determine unambiguously the Faraday rotation toward
the masing regions.

\section{Conclusions}
Modest levels of linear polarization are relatively common towards
strong 6.7~GHz methanol masers.  Comparison of the polarization
properties between the 6.7 and 12.2~GHz transition in one source
(NGC6334F) shows that for this case they are strikingly similar.  This
suggests that high spatial resolution polarization observations of the
two transitions may allow the determination of the intrinsic position
angle of the polarized emission, which in turn could be used to obtain
information on the line of sight magnetic field in these regions.


\begin{figure}
\plottwo{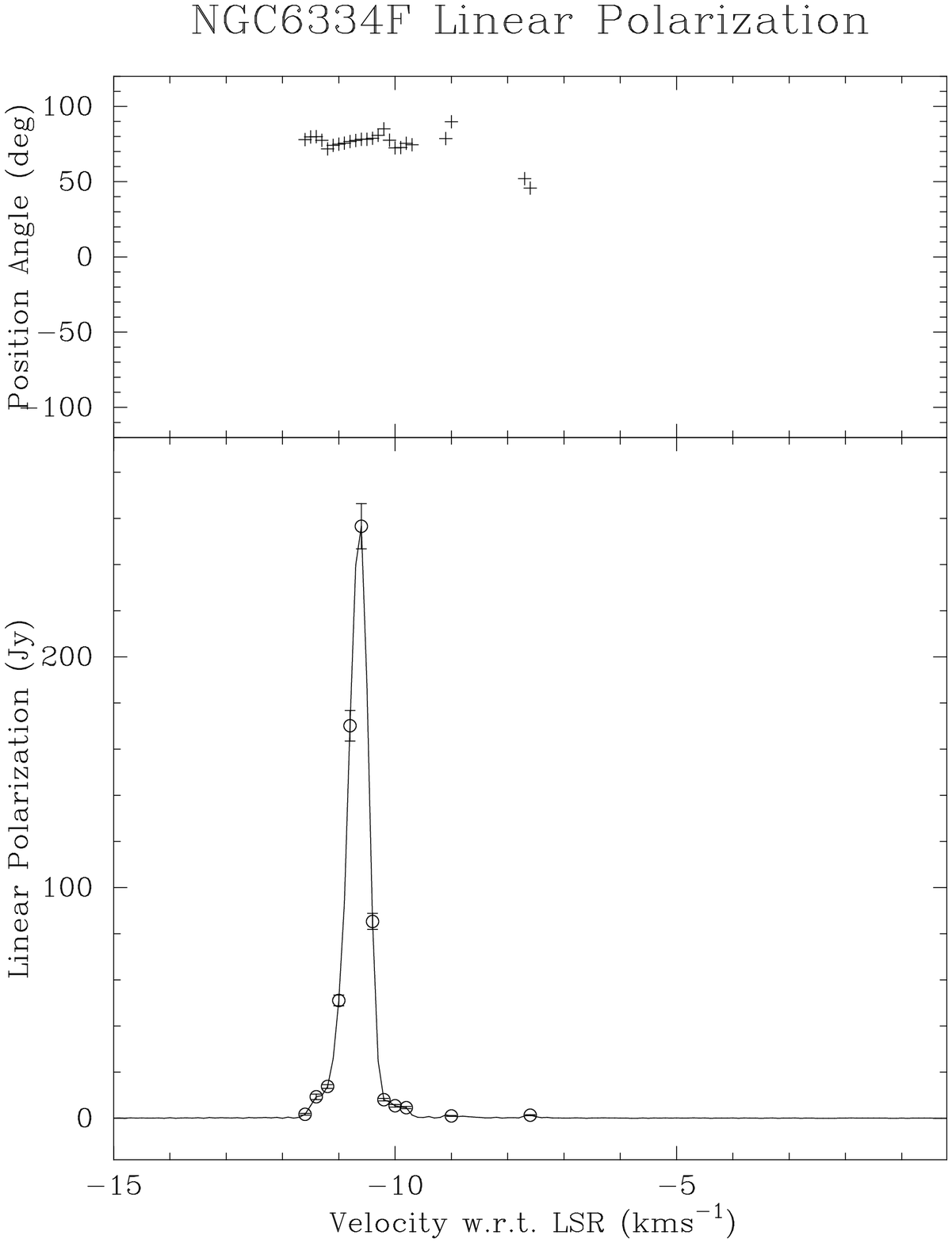}{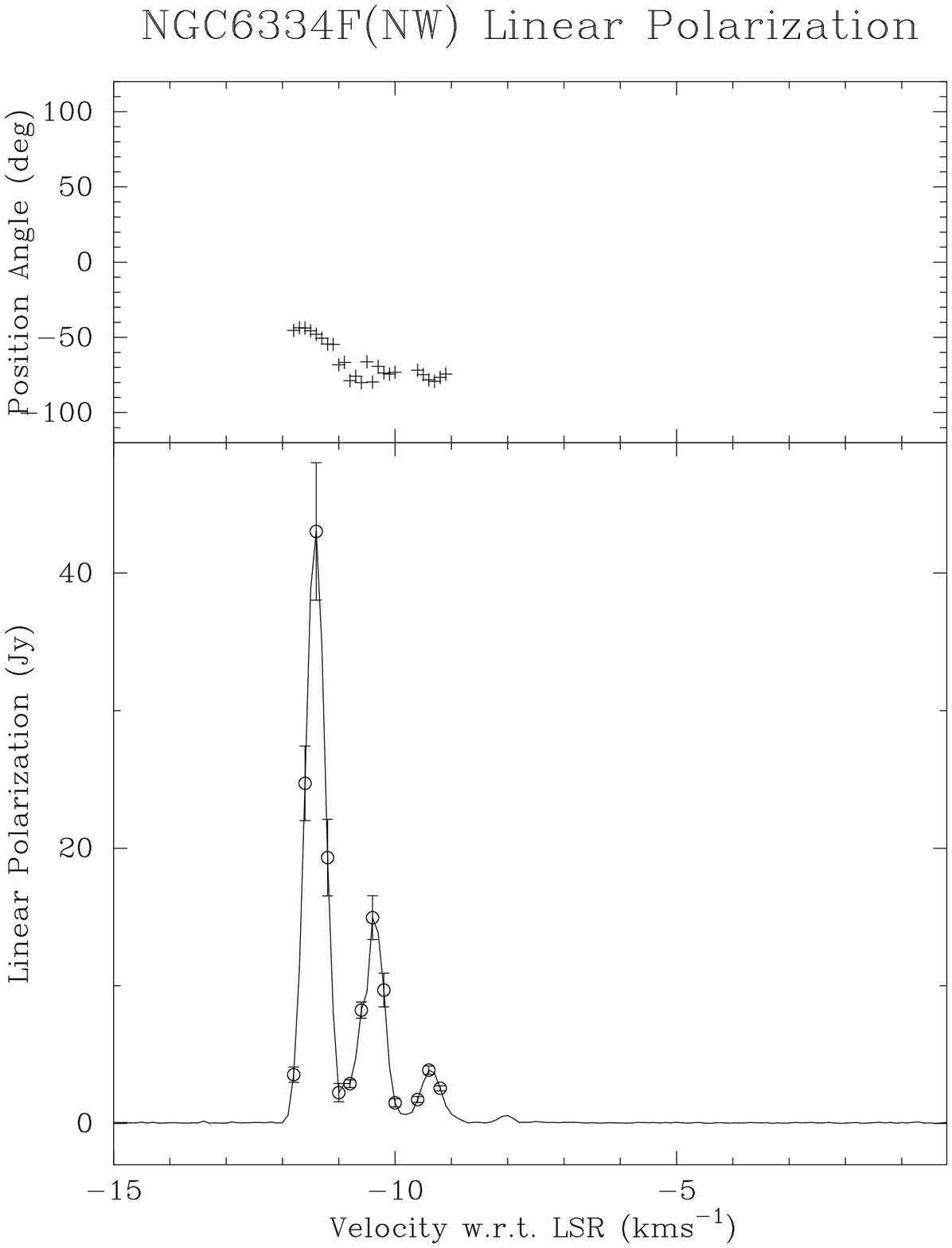}
\caption{The linear polarization spectrum of NGC6334F (left) and NGC6334F(NW)
  (right)}
\end{figure}

\end{document}